\documentclass[floats, eqnum, showpacs, nofootinbib, twocolumn, eqsecnum, preprintnumbers]{revtex4-1}
\usepackage[dvipdfmx]{graphicx}

\usepackage{color,graphicx,hyperref,dcolumn,amsmath}
\usepackage{amsfonts}
\usepackage{amssymb}

\def\b{\beta}

\def\E{{\cal E}}

\def\l{\ell}

\def\m{\mu}
\def\n{\nu}
\def\o{\omega}
\def\O{\Omega}
\def\pa{\partial}
\def\p{\phi}

\def\R{{\cal R}}
\def\s{\sigma}
\def\S{\Sigma}

\def\T{{\cal T}}
\def\ta{\tau}
\def\vp{\varphi}

\def\<{\langle}
\def\>{\rangle}
\def\={\equiv}

\begin{document}

\title{Effect of the self-gravity of shells on a high energy collision in a rotating Ba\~{n}ados-Teitelboim-Zanelli spacetime}

\author{Kota Ogasawara${}^{1}$}
\email{k.ogasawara@rikkyo.ac.jp}

\author{Naoki Tsukamoto${}^{2}$}
\email{tsukamoto@rikkyo.ac.jp}

\affiliation{${}^{1}$Department of Physics, Rikkyo University, Tokyo 171-8501, Japan}

\affiliation{${}^{2}$Frontier Research Institute for Interdisciplinary Sciences \& Department of Physics, Tohoku University, Sendai, Miyagi 980-8578, Japan}

\begin{abstract}
We consider a collision of two dust thin shells with a high center-of-mass (CM) energy including their self-gravity
in a Ba\~{n}ados-Teitelboim-Zanelli (BTZ) spacetime.
The shells divide the BTZ spacetime into three domains and the domains are matched by the Darmois-Israel junction conditions.
We treat only the collision of two shells which corotate with a background BTZ spacetime because of the junction conditions.
The counterpart of the corotating shell collision is a collision of two particles with vanishing angular momenta.
We compare the dust thin shell collision and the particle collision
in order to investigate the effects of the self-gravity of colliding objects on the high CM energy collision.
We show that the self-gravity of the shells affects the position of an event horizon and it covers the high-energy collisional event.
Therefore, we conclude that the self-gravity of colliding objects suppresses 
its CM energy and that any observer who stands outside of the event horizon cannot observe the collision with an arbitrary high CM energy.
\end{abstract}

\preprint{RUP-18-32}

\maketitle


\section{Introduction}
In 1977, Piran and Shaham~\cite{Piran:1977dm} 
discussed a collisional Penrose process which is a kind of an energy extraction~\cite{Penrose:1969pc} from a Kerr black hole 
and they found that the center-of-mass (CM) energy of two particles can be arbitrarily large in a near-horizon limit in the extremal Kerr spacetime.
In 2009, Ba\~{n}ados, Silk, and West (BSW) rediscovered the arbitrarily high CM energy of the particle collision 
and they pointed out that rotating black holes can act as particle accelerators~\cite{Banados:2009pr}.
The process is often called BSW collision or BSW process after BSW's work. 
See Harada and Kimura~\cite{Harada:2014vka} for a brief review on the BSW process.

In 2016, the LIGO Scientific and the Virgo Collaborations have reported the first detection of gravitational waves 
and they ensured the existence of astrophysical black holes~\cite{Abbott:2016blz}.
Physics in strong gravitational fields of black holes and compact objects such as BSW process
 would be more interest among researchers not only in general relativity
 but also in astronomy and astrophysics.

Several critical comments on the BSW process were given in Refs.~\cite{Berti:2009bk,Jacobson:2009zg,McWilliams:2012nx}.
It is well known that there is the upper bound of the angular momentum of the Kerr black hole in an astrophysical situation~\cite{Thorne:1974ve}.
It needs arbitrarily long proper time of either of two particles with the infinite CM energy
to reach the event horizon for a maximally rotating Kerr black hole.
If the self-gravity of the colliding particles is strong, the self-gravity will weaken the CM energy.
We should keep in mind that an observer distant from a black hole may not see
the products of the collision with high energy and/or very massive 
and the observed products must be highly red-sifted even if the CM energy is very large~\cite{McWilliams:2012nx}.

The details of the BSW collision have been investigated after stimulation by the criticism.
Patil~\textit{et al.} \cite{Patil:2015} considered a finite CM energy of a collision of two particles with a finite proper time.
The collision of particle with an innermost stable circular orbit~\cite{Harada:2010,Zaslavskii:2012ua},
off-equatorial-plane collisions~\cite{Harada:2011xz}, 
a collision in a weak electromagnetic field~\cite{Igata:2012js}, 
and the BSW collision in the near-horizon geometry~\cite{Galajinsky:2013as} 
have been investigated.
Nongeodesic particle collisions have been considered in~\cite{Tanatarov:2013gss,Tanatarov:2014fra}.
A close relation between the BSW collision and the Aretakis instability which is a test-field instability of an extremal horizon~\cite{Aretakis:2011ha,Aretakis:2011hc,Aretakis:2011gz,Aretakis:2012bm,Aretakis:2013dpa,Murata:2012ct,Murata:2013daa,Aretakis:2012ei} 
was pointed out~\cite{Tsukamoto:2013dna}.
The details of the collisional Penrose process
were also investigated by several authors~\cite{Bejger:2012yb,Harada:2012ap,Schnittman:2014zsa,Berti:2014lva,Leiderschneider:2015ika,Leiderschneider:2015kwa,Ogasawara:2015umo,Harada:2016eff,Ogasawara:2016yfk,Nakao:2018}. 

Particle collisions with high CM energy occur not only in a Kerr black hole spacetime 
but also in a Kerr naked singularity~\cite{Patil:2011ya}, 
a Kerr-Newmann~\cite{Wei:2010vca}, a Kerr-(anti-)de Sitter~\cite{Li:2010ej}, 
lower-dimensional~\cite{Lake:2010bq,Yang:2012we,Hussain:2012su,Sadeghi:2013gmf,Fernando:2017kut,Tsukamoto:2017rrl}, 
and higher-dimensional spacetimes~\cite{Abdujabbarov:2013qka,Tsukamoto:2013dna,Zaslavskii:2014xka,An:2017tlp}.
A particle collision with an unbounded CM energy in an extremal Reissner-Nordstr\"{o}m spacetime \cite{Zaslavskii:2010aw} 
and a higher-dimensional Reissner-Nordstr\"{o}m spacetime \cite{Tsukamoto:2013dna} have been investigated as the electromagnetic counterpart of the BSW collision.
\footnote{The collision of charged particles in a rotating and charged spacetime was also investigated by Hejda and Bi{\v c}{\' a}k \cite{Hejda:2017}.}
The BSW collision is regarded as a universal process in extremal spacetimes in all dimensions.

The self-gravity effect of colliding objects on the BSW collision cannot be neglected 
since the collision occurs in a near-horizon limit.
However, it is very difficult to treat the self-gravity of particles analytically in a Kerr spacetime.
Kimura~\textit{et al.} considered the collision of two thin shells~\cite{Lanczos:1922,Lanczos:1924} 
in a Reissner-Nordstr\"{o}m spacetime instead of a Kerr spacetime for its simplicity 
and they calculated analytically the CM energy of the shell collision including their self-gravity~\cite{Kimura:2010qy,Patil:2012,Nakao:2013uj}.

Can we treat thin shells in stationary and axisymmetric spacetimes?
The analytical treatment of a thin shell in the Kerr spacetime is very difficult~\cite{DeLaCruz:1968zz,Krasinski:1978,McManus:1991} 
but the difficulty of the technical problem depends on the dimension of the spacetime~\cite{Mann:2008rx,Crisostomo:2004,Delsate:2014iia,Rocha:2015tda}.
In Ref.~\cite{Mann:2008rx}, Mann~\textit{et al.} investigated the collapse of a shell in a three-dimensional stationary and axisymmetric spacetime and
they showed that the motion of the shell is tractable.

A black hole solution in three dimensions was obtained by Ba\~{n}ados, Teitelboim, and Zanelli (BTZ)~\cite{Banados:1992wn,Banados:1992gq}.
The BTZ spacetime has a negative cosmological constant 
since gravity in three dimensions is weaker than the one in four dimensions.
In the BTZ spacetime, particle motions \cite{Cruz:1994ir}, 
the BSW collision \cite{Lake:2010bq,Yang:2012we,Hussain:2012su,Tsukamoto:2017rrl}, 
gravitational perturbations induced by falling particles \cite{Rocha:2011wp}, 
and thermodynamics of thin shells~\cite{Lemos:2015zma,Lemos:2017mci,Lemos:2017aol} have been investigated.

In this paper, we investigate the collision of two dust thin shells 
in the BTZ spacetime with an angular momentum and a negative cosmological constant in three dimensions.
We use a thin-shell formalism
\cite{Darmois:1927,Israel:1966rt,Poisson:2004} 
for corotating thin shells investigated by Mann~\textit{et al.}~\cite{Mann:2008rx}. 
We study the collision of two particles with vanishing angular momenta as the counterpart of the shell collision
and then we investigate the effects of the self-gravity of the shells on their collision with a high CM energy.

The organization of this paper is as follows. 
In Sec. \ref{Sec:particle}, we investigate a collision of two particles with vanishing angular momenta.
We review the thin shell formalism in a corotating frame in Sec. \ref{Sec:shell}.
In Sec. \ref{Sec:shell collision}, we investigate the collision of two dust thin shells.
Section \ref{Sec:Discussion and Conclusion} is devoted to the discussion and conclusion.
In this paper, we use the units in which the speed of light and $8G$ are unity as in Sec. III of Ref. \cite{Mann:2008rx}, 
where $G$ is Newton's constant in three dimensions.


\section{Particle collision in the BTZ spacetime}
\label{Sec:particle}
In this section, we review the center-of-mass energy of the collision of two particles in the BTZ spacetime.
The metric
\cite{Banados:1992wn,Banados:1992gq}
is given by
\begin{equation}\label{eq:line_element1}
ds^2=-f(r)dt^2+\frac{dr^2}{f(r)}+r^2 \left[d\vp- \O(r) dt \right]^2,
\end{equation}
where
\begin{eqnarray}
f(r)&\=&-M+\frac{r^2}{\l^2}+\frac{J^2}{4r^2},\\
\O(r)&\=&-\frac{g_{t\vp}}{g_{\vp\vp}}=\frac{J}{2r^2},\\
\l&\=&\sqrt{\frac{1}{-\Lambda}}.
\end{eqnarray}
Here $M$, $J$, and $\O(r)$ are the mass, the angular momentum, and angular velocity of the spacetime, respectively, 
and $\l$ is the scale of a curvature related to the negative cosmological constant $\Lambda<0$.
The spacetime is a stationary and axisymmetric spacetime with two Killing vectors $\pa_t$ and $\pa_\vp$.
Without loss of generality, we assume that the angular momentum $J$ is non-negative.
In this paper, we concentrate on the case where $M$ is positive.
The spacetime has an event horizon at
\begin{equation}
r=r^H\=\l\sqrt{ \frac{M}{2} \left( 1+ \sqrt{1-\frac{J^2}{\l^2M^2}} \right) },
\end{equation}
for $J\leq\l M$
and it is called extremal spacetime for $J=\l M$.
For $J>\l M$, it has a naked singularity. 
We discuss the last case in Appendix~\ref{App:over-spinning_particle}.

We consider a particle motion with a 3-momentum $p^\m$ and a rest mass $m$.
The conserved energy and angular momentum of the particle are given by
\begin{eqnarray}
E\=-g_{\m\n}(\pa_t)^\m p^\n
\;\;{\rm and}\;\;
L\=g_{\m\n}(\pa_\vp)^\m p^\n,
\label{EL}
\end{eqnarray}
respectively.
From Eq. (\ref{EL}) and the condition $p^\m p_\m=-m^2$, we obtain the components of the 3-momentum as
\begin{eqnarray}
p^t(r)&=&\frac{S(r)}{f(r)},
\label{pt:particle}\\
p^r(r)&=&\s\sqrt{R(r)},
\label{pr:particle}\\
p^\vp(r)&=&\frac{\O(r)S(r)}{f(r)}+\frac{L}{r^2},
\label{pvp:particle}
\end{eqnarray}
where $\s$, $S(r)$, and $R(r)$ are defined as
\begin{eqnarray}
\s&\=&{\rm sgn}(p^r)=\pm1,\\
S(r)&\=&E-\O(r)L,\label{def:S}\\
R(r)&\=&S^2(r)-\left(m^2+\frac{L^2}{r^2}\right)f(r),\label{def:R}
\end{eqnarray}
respectively.
Note that a forward-in-time condition
\begin{eqnarray}
p^t(r)\geq0,
\end{eqnarray}
must be satisfied for a particle motion.

An energy equation which describes the radial motion of the particle is given by, 
from Eq.~(\ref{pr:particle}), 
\begin{eqnarray}
\left(\frac{dr}{d\ta}\right)^2+V(r)=0,
\end{eqnarray}
where $V(r)\= -R(r)$ is the effective potential for the radial motion of the particle and $\ta$ is its proper time.
Here we have used relations $p^\m=mu^\m$ and $u^\m=dx^\m/d\ta$, where $u^\m$ is the 3-velocity of the particle.
We call a condition $V(r^H)=0$ critical condition.
The critical condition is rewritten as
\begin{eqnarray}
E-\O_HL=0,
\label{def:critical}
\end{eqnarray}
where
\begin{eqnarray}
\O_H\=\O(r^H)=\frac{J}{2(r^H)^2},
\end{eqnarray}
is the angular velocity of the horizon.

The angular velocity of a particle $\o(r)$ is defined as
\begin{eqnarray}
\o(r)\=\frac{d\vp}{dt}=\frac{p^\vp}{p^t}.
\label{def:omega}
\end{eqnarray}
When a particle has a zero conserved angular momentum $L=0$, from 
Eqs. (\ref{pt:particle}), (\ref{pvp:particle}), and (\ref{def:omega}), 
the angular velocity of the particle coincides with the angular velocity of the spacetime, i.e., $\o(r)=\O(r)$.
This means that the particle with $L=0$ corotates with the background spacetime.

We concentrate on the motion of a particle with $L=0$ which 
is the counterpart of a shell corotating with the background spacetime.
Using a dimensionless radial coordinate $x\=r/\l$, 
the effective potential of the particle with the specific energy $e\=E/m$ and the position of the event horizon are expressed as
\begin{eqnarray}
V(x)&=&-e^2+f(x) \nonumber\\
&=&x^2-M-e^2+\frac{c}{x^2},
\label{V_particle}
\end{eqnarray}
and 
\begin{eqnarray}
x^H\=\frac{r^H}{\l}=\sqrt{\frac{M+\sqrt{M^2-4c}}{2}},
\end{eqnarray}
respectively, 
where 
\begin{eqnarray}
f(x)=x^2-M+\frac{c}{x^2},\;\;
c\=\frac{J^2}{4\l^2}.
\label{f_particle}
\end{eqnarray}
As $x$ increases from $0$ to infinity, $V(x)$ and $f(x)$ begin with infinity, monotonically decrease to a local minimum at $x=x^m\=c^{\frac{1}{4}}$, and monotonically increase to infinity.
The particle motion is restricted to a region $x^-\leq x\leq x^+$ where the effective potential $V(x)$ is nonpositive.
The boundaries $x^\pm$ are obtained as
\begin{eqnarray}
x^\pm\=\sqrt{\frac{M+e^2\pm\sqrt{(M+e^2)^2-4c}}{2}}.
\label{xpm_particle}
\end{eqnarray}
We notice a relation
\begin{eqnarray}\label{eq:inequality}
x^-\leq x^m\leq x^H\leq x^+.
\end{eqnarray}
We obtain $x^m= x^H$ in the extremal case, $x^H= x^+$ in the critical case, 
and $x^-= x^m= x^H= x^+$ in the extremal and critical case.

We consider a collision of two particles, named particle $1$ and $2$, with vanishing conserved angular momenta $L_1=L_2=0$
in a region $x^H\leq x\leq x^+_a$ where is seen by an observer who stands at the outside of the horizon $x^H$.
Hereinafter $p^\m_a$, $E_a$ $e_a$, $m_a$, $x^\pm_a$, $V_a$, 
and $\s_a$ denote $p^\m$, $E$, $e$, $m$, $x^\pm$, $V$, and $\s$ of particle $a$~($a=1$ and $2$), respectively.
From Eq.~(\ref{def:critical}), the critical condition for particle $a$ with $L_a=0$ is given by
\begin{eqnarray}
E_a=0
,\;\;{\rm i.e.,}\;\;
e_a=0.
\label{critical:particle}
\end{eqnarray}
The CM energy $E_{\rm cm}(x)$ of the particles is given by
\begin{eqnarray}
E^2_{\rm cm}(x)&\=&-g_{\m\n}\left(m_1u^\m_1+m_2u^\m_2\right)\left(m_1u^\n_1+m_2u^\n_2\right)\nonumber\\
&=&m^2_1+m^2_2+2m_1m_2\frac{e_1e_2-\s_1\s_2\sqrt{V_1(x)V_2(x)}}{f(x)}.
\nonumber\\
\end{eqnarray}
Let us consider a rear-end collision, i.e., we choose $\s_1=\s_2=-1$.
As $x$ increases from $x^H$ to $x^+_a$, the CM energy begins with 
\begin{eqnarray}
E_{\rm cm}(x^H)=\sqrt{m^2_1+m^2_2+m_1m_2\left(\frac{e_2}{e_1}+\frac{e_1}{e_2}\right)},
\end{eqnarray}
and monotonically increases to
\begin{equation}
E_{\rm cm}(x^+_a)=E_{\rm cm}^{\rm max},
\label{Ecmx^+}
\end{equation}
where
\begin{eqnarray}
E_{\rm cm}^{\rm max}
\= \sqrt{m^2_1+m^2_2+2m_1m_2\frac{e_1e_2}{e^2_a}}. 
\label{Ecmmaxparticle}
\end{eqnarray}
We have used l'Hopital's rule to estimate
$E_{\rm cm}(x^H)$.
We get $E_{\rm cm}(x)=m_1+m_2$ if the both particles have a same specific energy $e_1=e_2$.
We set particle~$1$ to be inner than particle~$2$ in the rest of this section.

We are interested in the collision of the inner particle with a critical limit $e_1\to0$ and the outer particle 
which is not critical in the extremal BTZ spacetime
since the collision will correspond with the BSW collision 
in the extremal Kerr spacetime~\cite{Banados:2009pr} and in the extremal Reissner-Nordstr\"om spacetime~\cite{Zaslavskii:2010aw}. 
In this case, the collisional point must be $x \to x^H +0$ because of inequality~(\ref{eq:inequality}) and $x^{\pm}_1 \to x^H \pm 0$ 
and the CM energy $E_{\rm cm}$ diverges there. 
Notice that a particle with the critical condition has 
$V_1(x^H)=V_{1}^{'}(x^H)=0$
and $V_{1}^{''}(x^H)>0$,
where a prime denotes a derivative with respect to $x$,
in the extremal BTZ spacetime
while one has $V_1(x^H)=V_{1}^{'}(x^H)=0$ and $V_{1}^{''}(x^H)<0$ 
in an extremal Kerr black hole spacetime \cite{Banados:2009pr} and an extremal Reissner-Nordstr\"om spacetime~\cite{Zaslavskii:2010aw}.
The positive sign of $V_{1}^{''}(x^H)$ is caused by the negative cosmological constant and it will not affect on the BSW collision strongly 
as long as the collision occurs near the horizon.


\section{Dust thin shell and its motion in the BTZ spacetime}
\label{Sec:shell}
In this section, as a preparation to study a shell collision, 
we review the Darmois-Israel junction conditions \cite{Darmois:1927,Israel:1966rt,Poisson:2004} 
and the motion of a dust thin shell in the BTZ spacetime \cite{Mann:2008rx}.
Hereinafter, we use $x^\m$ denoting coordinates in every domain for simplicity. 

\begin{figure}[t]
\begin{center}
\includegraphics[width=60mm]{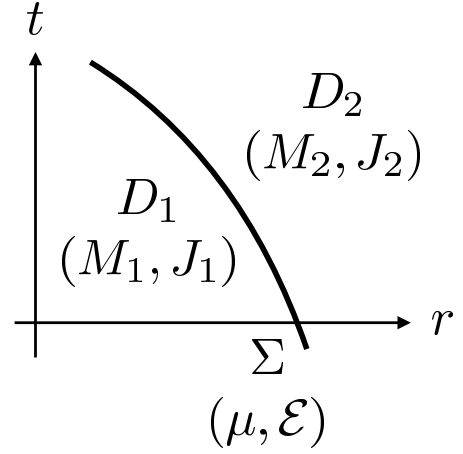}
\caption{Schematic picture of the BTZ spacetime divided into two domains $D_1$ and $D_2$ by the hypersurface $\Sigma$.
Domain~$D_A$ ($A=1$ and $2$) has a mass~$M_A$ and an angular momentum~$J_A$.
A thin shell on the hypersurface $\Sigma$ has a proper mass $\mu$ and specific energy ${\cal E}\equiv(M_2-M_1)/\mu$. 
}
\label{fig:Shell1}
\end{center}
\end{figure}

We consider a two-dimensional hypersurface $\S$ which divides the BTZ spacetime into an interior domain $D_1$ and an exterior domain $D_2$.
Domain $D_A$ ($A=1$ and $2$) has a mass~$M_A$ and an angular momentum~$J_A$, see Fig.~\ref{fig:Shell1}.
For simplicity, we assume that $D_1$ and $D_2$ have the same $\l$.
We assume that we can take same coordinates $y^i$ on $\S$ in both the domains.

The projection operator from the three-dimensional BTZ spacetime to $\S$ is defined as
\begin{eqnarray}
e^\m_i\=\frac{\pa x^\m}{\pa y^i}.
\end{eqnarray}
The induced metric on the hypersurface~$\S$ in domain~$D_A$ is defined as
\begin{eqnarray}
g^{A}_{ij}\=g^A_{\m\n}e^\m_ie^\n_j.
\end{eqnarray}
Using a unit vector~$n^\m$ normal to the hypersurface~$\S$, which directed from $D_1$ to $D_2$,
the extrinsic curvature of the hypersurface~$\S$ in the domain~$D_A$
is defined as
\begin{eqnarray}
K^A_{ij}\=e^\m_ie^\n_j\nabla^A_\m n_\n,
\label{def:Kij}
\end{eqnarray}
where $\nabla^A_\m$ is the covariant derivative within $D_A$.
The first and second junction conditions are given by
\begin{eqnarray}
[g_{ij}]=0
\;\;{\rm and}\;\;
[K_{ij}]=0,
\end{eqnarray}
respectively.
Here the bracket is defined as
\begin{eqnarray}
[\Psi]\=\Psi(D_2)|_\S-\Psi(D_1)|_\S,
\end{eqnarray}
where $\Psi$ is any quantity defined on the both sides of $\S$.

When we introduce a corotating frame on $\S$ with an azimuth coordinate 
\begin{eqnarray}
d\phi\=d\varphi-\frac{J_A}{2\R^2(t)}dt,
\end{eqnarray}
the metric in $D_A$ is given by
\begin{eqnarray}
ds^2_A&=&g^A_{\m\n}dx^\m dx^\n\nonumber\\
&=&-f_A(r)dt^2+\frac{dr^2}{f_A(r)}\nonumber\\
&&+r^2\left[d\phi+\frac{J_A}{2}\left(\frac{1}{\R^2(t)}-\frac{1}{r^2}\right)dt\right]^2,
\end{eqnarray}
where
\begin{eqnarray}
f_A(r)\=-M_A+\frac{r^2}{\l^2}+\frac{J_A^2}{4r^2}.
\label{eq:f_a}
\end{eqnarray}

From the first junction condition $[g_{ij}]=0$, the induced metric~$g_{ij}^{A}$ on $\S$ with
\begin{eqnarray}
t=\T(\ta)
\;\;{\rm and}\;\;
r=\R(\T(\ta))=\R(\ta),
\label{tTrR}
\end{eqnarray}
is given by
\begin{eqnarray}
ds^2_\S&=&g_{ij}^{A}dy^idy^j=-d\ta^2+\R^2(\ta)d\phi^2\nonumber\\
&=&
\left[-f_A(\R)\dot{\T}^2+\frac{\dot{\R}^2}{f_A(\R)}\right]d\ta^2+\R^2(\ta)d\phi^2,
\label{gij}
\nonumber\\
\end{eqnarray}
where a dot denotes a derivative with respect to the proper time $\ta$ of an observer on $\S$.
This implies
\begin{eqnarray}
f_A(\R)\dot{\T}=\b_A,
\label{beta}
\end{eqnarray}
where 
\begin{eqnarray}
\b_A\=\sqrt{\dot{\R}^2+f_A(\R)}.
\label{betadef}
\end{eqnarray}

The components of $e^\m_i$ and $n_\m$ are described by
\begin{eqnarray}
e^\m_\ta=\left(\dot{\T},\dot{\R},0\right),\;\;
e^\m_\p=(0,0,1),
\end{eqnarray}
and
\begin{eqnarray}
n_\m=\left(-\dot{\R},\dot{\T},0\right),
\end{eqnarray}
respectively.
We obtain the components of the extrinsic curvature~(\ref{def:Kij}) and its trace $K^A\=g^{ij}K^A_{ij}$ as
\begin{eqnarray}
K^A_{\ta\ta}=-\frac{\dot{\beta_A}}{\dot{\mathcal{R}}},\;\;
K^A_{\phi \phi}=\R \beta_A,\;\;
K^A_{\tau \phi}=\frac{J_A}{2\mathcal{R}},
\label{Kij}
\end{eqnarray}
and
\begin{eqnarray}
K^A=\frac{\dot{\beta_A}}{\dot{\mathcal{R}}}+\frac{\beta_A}{\mathcal{R}},
\label{K}
\end{eqnarray}
respectively.
We notice that the second junction condition $[K_{ij}]=0$ is violated unless we consider a trivial case.
Therefore, we introduce a thin shell on $\S$ following equations 
\begin{eqnarray}
\pi S_{ij}=-\left([K_{ij}]-[K]g_{ij}\right),
\label{S=K}
\end{eqnarray}
where $S_{ij}$ is the surface stress-energy tensor of the thin shell.

Let us consider a dust thin shell with 
the surface stress-energy tensor $S_{ij}$ given by
\begin{eqnarray}
S_{ij}=\rho u_iu_j,
\label{Sij}
\end{eqnarray}
where $\rho$ and $u_i=(-1,0)$ are the surface energy density and the 2-velocity of the shell, respectively.
Using Eqs.~(\ref{gij}), (\ref{Kij}), (\ref{K}), and (\ref{Sij}), the $(\tau, \tau)$, $(\phi, \phi)$, and $(\tau, \phi)$ components of Eq. (\ref{S=K}) 
are obtained as
\begin{eqnarray}
[\b]+\pi\rho\R&=&0,\label{JCtt}\\
~[\dot{\b}]&=&0,\label{JCpp}
\end{eqnarray}
and
\begin{eqnarray}
[J]=0,\label{JCtp}
\end{eqnarray}
respectively.
From Eqs.~(\ref{JCtt}) and (\ref{JCpp}), we obtain
\begin{eqnarray}
\frac{d}{d\ta}(\pi\rho\R)=0.
\end{eqnarray}
Therefore, we can define the proper mass $\mu$ of the shell
\begin{eqnarray}
\m\=2\pi\rho\R,
\label{def:mu}
\end{eqnarray}
which is constant along its trajectory.
We define the specific energy $\E$ of the shell as
\begin{eqnarray}
\E\=\frac{[M]}{\m}.
\label{def:E}
\end{eqnarray}
We assume that the proper mass $\m$ and the specific energy $\E$ are positive.
This implies that the masses satisfy the relation $M_1<M_2$.
From Eq. (\ref{JCtp}), the angular momenta in all the domains must be the same, i.e.,
\begin{eqnarray}
J\=J_1=J_2.
\label{same_J}
\end{eqnarray}

Using Eqs. (\ref{betadef}), (\ref{JCtt}), and (\ref{def:mu}), we obtain the energy equation as
\begin{eqnarray}
\left(\frac{d\R}{d\ta}\right)^2+V(\R)=0,
\end{eqnarray}
where $V(\R)$ is the effective potential of the shell motion.
Using $x\=\R/\l$, the effective potential is expressed as
\begin{eqnarray}
V(x)=x^2-\<M\>-\E^2+\frac{c}{x^2}-\frac{\m^2}{16},
\label{V_shell}
\end{eqnarray}
where
\begin{eqnarray}
\<M\>\=\frac{M_2+M_1}{2}.
\end{eqnarray}
As $x$ increases from $0$ to infinity, $V(x)$ begins with infinity, monotonically decreases to a local minimum at $x=x^m\=c^{\frac{1}{4}}$, and monotonically increases to infinity.

The shell motion is restricted to the region $x^-\leq x\leq x^+$ where the effective potential is nonpositive.
Here 
\begin{eqnarray}
x^\pm\=\sqrt{\frac{b\pm\sqrt{b^2-4c}}{2}},
\end{eqnarray}
have been obtained as the positive solutions of $V(x)=0$, 
where
\begin{eqnarray}
b\=\frac{\m^2}{16}+\<M\>+\E^2
=\left(\frac{\m}{4}-\E\right)^2+M_2.
\end{eqnarray}

We will call a limit
$\m\to0$ and $M_1 \rightarrow {\cal M} \= M_2$ with $\E=(\rm{constant})\neq0$
test shell limit.
In the test shell limit, the effective potential of the shell (\ref{V_shell}) is obtained as
\begin{eqnarray} 
V(x)=x^2-{\cal M}-\E^2+\frac{c}{x^2},
\label{V_shell3}
\end{eqnarray}
and it takes the same form as the effective potential of a particle (\ref{V_particle}).


\section{Dust Thin Shell Collision in the BTZ spacetime}
\label{Sec:shell collision}
\begin{figure}[t]
\begin{center}
\includegraphics[width=80mm]{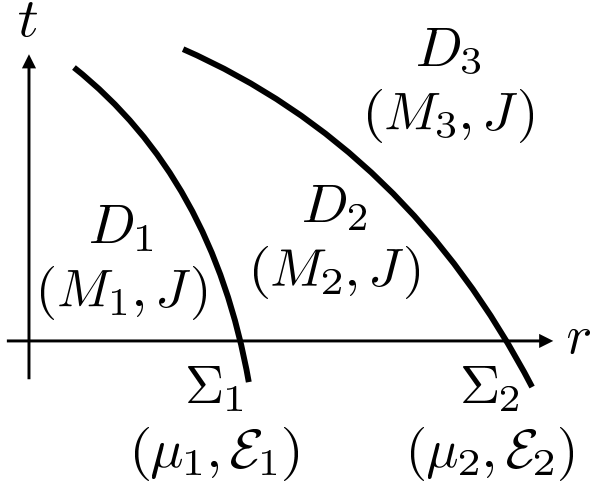}
\caption{Schematic picture of the BTZ spacetime divided into three domains $D_1$, $D_2$, and $D_3$ by hypersurfaces $\S_1$ and $\S_2$.
Domain $D_A$ ($A=1$, $2$, and $3$) is the BTZ spacetime with a mass $M_A$ and the same angular momentum $J$.
A thin shell $a$ ($a=1$ and $2$) on the hypersurfaces $\Sigma_a$ has a proper mass $\mu_a$ and specific energy ${\cal E}_a\equiv(M_{a+1}-M_a)/\mu_a$.
} 
\label{fig:Shell2}
\end{center}
\end{figure}

In this section, we investigate the collision of two dust thin 
shells in the BTZ spacetime.
We assume that shell 1 and shell 2 are on an inner hypersurface $\S_1$ and an outer hypersurface $\S_2$, respectively.
These two hypersurfaces divide the BTZ spacetime into an interior domain $D_1$, a middle domain $D_2$, and an exterior domain $D_3$.
We assume that every domain~$D_A$ ($A=1$, $2$, and $3$) is the BTZ spacetime with the same $\l$ for simplicity.
From Eq. (\ref{JCtp}), all the domains have the same angular momenta $J$. 
 See Fig.~\ref{fig:Shell2}.

When $J\leq\l M_A$, where $M_A$ is a mass in $D_A$, is satisfied, 
the position of the event horizon is obtained as $x=x^H_A$, 
where
\begin{eqnarray}
x^H_A\=\sqrt{\frac{M_A+\sqrt{M_A^2-4c}}{2}}.
\label{horizon}
\end{eqnarray}
We assume $M_1<M_2<M_3$.
We consider five cases according to the value of $J$ as shown in Table~\ref{5cases}:
case I (II) for $J<\l M_1$ ($J=\l M_1$),  case III for $J=\l M_2$, and case IV (V) for $J=\l M_3$ ($J>\l M_3$).

\begin{table}[h]
\begin{center}
\begin{tabular}{|c||c|c|c|c|c|}
\hline
~case~ & ~$D_1$~ & ~$D_2$~ & ~$D_3$~ & $J$ & $X$ \\ \hline\hline
I & S & S & S & ~$J<\l M_1$~ & ~$X<0$~ \\ \hline
II & E & S & S & $J=\l M_1$ & $X<0$ \\ \hline
III & O & E & S & $J=\l M_2$ & $X=0$ \\ \hline
IV & O & O & E & $J=\l M_3$ & $X>0$ \\ \hline
V & O & O & O & $J>\l M_3$ & $X>0$ \\
\hline
\end{tabular}
\caption{Five cases of the spacetime according to the value of $J$.
Symbols S, E, and O denote the subextremal, extremal, and overspinning spacetimes, respectively.
$X$ is defined as the minimum value of $f_2(x)$.}
\label{5cases}
\end{center}
\end{table}

The effective potential of shell $a$~($a=1$ and $2$) is given by
\begin{eqnarray}
V_a(x)=-Z^2_a+f(x),
\label{V_a}
\end{eqnarray}
where
\begin{eqnarray}
Z_1&\=&\frac{\m_1}{4}-\E_1,\\
Z_2&\=&\frac{\m_2}{4}+\E_2,\\
f(x)&\=&f_2(x)=x^2-M_2+\frac{c}{x^2}.
\end{eqnarray}
As $x$ increases from $0$ to infinity, $V_a(x)$ begins with infinity, monotonically decreases to
a local minimum
 \begin{eqnarray}
V_a(x^m)=-Z^2_a+X,
\end{eqnarray}
where
\begin{eqnarray}
X\=f(x^m)=2\sqrt{c}-M_2,
\label{def:X}
\end{eqnarray}
at $x=x^m\=c^{\frac{1}{4}}$, and monotonically increases to infinity.
Note that $x^m$ for shells $1$ and $2$ are the same.

A motion of shell $a$ is restricted to the region $x^-_a\leq x\leq x^+_a$,
where the effective potential $V_a(x)$ is nonpositive.
Here $x^\pm_a$ is given by 
\begin{eqnarray}
x^\pm_a\=\sqrt{\frac{b_a\pm\sqrt{b^2_a-4c}}{2}},
\label{x^pm}
\end{eqnarray}
and
\begin{eqnarray}
b_a\=Z_a^2+M_2.
\end{eqnarray}
In order for such the region to exist, a condition
\begin{eqnarray}
X\leq Z^2_a,
\label{min:Z}
\end{eqnarray}
must be satisfied.

For a comparison with a particle satisfying the critical condition~(\ref{critical:particle}),   
one might have an interest in a shell with~$V_1(x_2^H)=V_1'(x_2^H)=0$.
We can show easily that 
\begin{eqnarray}
V_1(x^H_2)=V'_1(x^H_2)=0,
\end{eqnarray}
is satisfied if and only if 
\begin{eqnarray}
Z_1=0,
\label{critical:shell}
\end{eqnarray}
and $J=\l M_2$ are satisfied.
Therefore, we call the condition $Z_1=0$ critical condition for shell~$1$

\subsection{Center-of-mass energy}
The CM energy $E_{\rm cm}(x)$ of two shells at a collisional point
is given by \cite{Kimura:2010qy,Nakao:2013uj}
\begin{eqnarray}
E^2_{\rm cm}(x)&\=&
-g_{\m\n}\left(\m_1U_1^\m+\m_2U_2^\m\right)\left(\m_1U_1^\n+\m_2U_2^\n\right)
\nonumber\\&=&
\m_1^2+\m_2^2+2\m_1\m_2
\left(f(x)\dot{\T}_1\dot{\T}_2-\frac{\dot{\R}_1\dot{\R}_2}{f(x)}\right),
\nonumber\\
\label{def:Ecmshell}
\end{eqnarray}
where $g_{\m\n}$ is the metric in $D_2$, $U_a^\m=(\dot{\T}_a,\dot{\R}_a,0)$ is the 3-velocity of shell~$a$,
where
\begin{eqnarray}
\dot{\T}_a&\=&\frac{\sqrt{f(x)-V_a(x)}}{f(x)},\label{dotTa}\\
\dot{\R}_a&\=&\s_a\sqrt{-V_a(x)},\label{dotRa}
\end{eqnarray}
and $\s_a\={\rm sgn}(\dot{\R}_a)=\pm1$.

We will concentrate on the rear-end collision of two dust thin shells.
In this case, we should choose $\s_1=\s_2=-1$ and the CM energy is expresses as
\begin{eqnarray}
E^2_{\rm cm}(x)=\m^2_1+\m^2_2+2\m_1\m_2\frac{|Z_1|Z_2-\sqrt{V_1(x)V_2(x)}}{f(x)}.
\nonumber\\
\end{eqnarray}
We notice $E_{\rm cm}(x)=\m_1+\m_2$ when $|Z_1|=Z_2$ is satisfied. 
The shells must satisfy a relation $\dot{\R}_1(x) \leq  \dot{\R}_2(x)$ to collide at the point.
It implies $V_1(x) \geq V_2(x)$, $\left| Z_1 \right| \leq Z_2$, or $x^+_1\leq x^+_2$.

\subsubsection{Cases I, II, and III}
In the cases I, II, and III, 
we concentrate on the region $x^H_2\leq x\leq x^+_1$ as in particle collisions.
As $x$ increases from $x^H_2$ to $x^+_1$, the CM energy begins with  
\begin{eqnarray}
E_{\rm cm}(x^H_2)=\sqrt{\m^2_1+\m^2_2+\m_1\m_2\left(\frac{Z_2}{|Z_1|}+\frac{|Z_1|}{Z_2}\right)},~
\end{eqnarray}
and monotonically increases to 
\begin{eqnarray}
E_{\rm cm}(x^+_1) = E^{\rm max}_{\rm cm},
\label{Ecmx^+_1}
\end{eqnarray}
where
\begin{eqnarray}
E^{\rm max}_{\rm cm} \= \sqrt{\m^2_1+\m^2_2+2\m_1\m_2\frac{Z_2}{|Z_1|}}.
\label{Ecmx^+_1b}
\end{eqnarray}
Here we have used l'Hopital's rule to calculate $E_{\rm cm}(x^H_2)$.
In a critical limit $Z_1\to0$, both $E_{\rm cm}(x^H_2)$ and $E_{\rm cm}(x^+_1)$ are arbitrarily large 
and $x^+_1$ coincides with $x^H_2$.
The event horizon, however, moves from $x^H_2$ to $x^H_3$ because of the self-gravity of shell~2.
Thus, the arbitrarily large CM energy cannot be seen by an observer outside the event horizon $x^H_3$.

\subsubsection{Cases IV and V}
In cases~IV and V, $x^H_2$ does not exist. 
Let us consider the region $x^-_1\leq x\leq x^+_1$.
The CM energy monotonically decreases from
\begin{eqnarray}
E_{\rm cm}(x^-_1)=E^{\rm max}_{\rm cm}, 
\end{eqnarray}
to 
\begin{eqnarray}
&&E_{\rm cm}(x^m)= \nonumber\\
&& \sqrt{\m^2_1+\m^2_2+2\m_1\m_2 \frac{|Z_1|Z_2-\sqrt{(Z^2_1-X)(Z^2_2-X)}}{X}}, \nonumber\\
\end{eqnarray}
as $x$ increases from $x^-_1$ to $x^m$
and it monotonically increases as from $x^m$ to $x^+_1$ 
and then it reaches
\begin{eqnarray}
E_{\rm cm}(x^+_1)=E^{\rm max}_{\rm cm}.
\label{Ecmx^pm_1}
\end{eqnarray}

In these cases, a critical shell with $Z_1=0$ is forbidden since its effective potential is positive.
When $V_1(x^m)=0$, i.e., $\left| Z_1 \right| = \sqrt{X}$, 
shell~$1$ can be only at $x=x^m=x^\pm_1$.
The CM energy there is given by 
\begin{eqnarray}
E_{\rm cm}(x=x^m=x^\pm_1)=\sqrt{\m^2_1+\m^2_2+\frac{2\m_1\m_2Z_2}{\sqrt{X}}}.
\end{eqnarray}


\subsection{Collision at $x \geq  x^H_3$}
Let us consider a shell collision at $x \geq  x^H_3$ for the cases I-IV. 
An observer at $x \geq  x^H_3$ in domain $D_3$ may see the parts of the products after the collision.
A condition $x^H_3\leq x^+_1$ must be satisfied for the existence of the inner shell~$1$.
From Eqs.~(\ref{horizon}) and (\ref{x^pm}), the condition $x^H_3\leq x^+_1$ is expressed as
\begin{eqnarray}
\m_2\E_2\leq Z^2_1.
\label{must:observable}
\end{eqnarray}
The finite upperbound of the CM energy is given by, from Eqs. (\ref{Ecmx^+_1}), (\ref{Ecmx^pm_1}), and (\ref{must:observable}),
\begin{eqnarray}
E_{\rm cm}(x=x^H_3=x^+_1)=\sqrt{\m^2_1+\m^2_2+\frac{\m_1\sqrt{\m_2}(\m_2+4\E_2)}{2\sqrt{\E_2}}}.\nonumber\\
\label{Ecm_max}
\end{eqnarray}
This shows that the self-gravity caused by the colliding shells suppresses the CM energy.
For an equal mass~$\m \= \m_1 =\m_2$, the upperbound of the CM energy is given by 
\begin{eqnarray}
E_{\rm cm}(x=x^H_3=x^+_1) = \frac{\mu^\frac{3}{4}}{\sqrt{2}\E_2^\frac{1}{4}}\left( 2\sqrt{\E_2}+\sqrt{\mu} \right),~
\end{eqnarray}
and it becomes, for a small mass $\m \ll \E_2$,   
\begin{eqnarray}
E_{\rm cm}(x=x^H_3=x^+_1) \simeq 2^{1/2}\E_2^{1/4}\m^{3/4}.
\label{Ecm_max2}
\end{eqnarray}


\subsection{Test shell limit}
Here we consider test shell limits for shells~$1$ and $2$
$\m_1\to0$ and $M_1 \to {\cal M}_- \equiv M_2$ with $\E_1=(\rm{constant})\neq0$
and
$\m_2\to0$ and $M_2 \to {\cal M}_+ \equiv M_3$ with $\E_2=(\rm{constant})\neq0$,
respectively.
Notice $x_a^H \rightarrow x_{a+1}^H$ in the test shell limit for shell~$a$. 
From Eqs.~(\ref{Ecmx^+_1}) and (\ref{Ecmx^+_1b}), the CM energy of the shells with the equal mass $\mu$ 
at $x=x^+_1$ in the test shell limits for shells~$1$ and $2$ is obtained as
\begin{eqnarray}
\frac{E^2_{\rm cm}(x^+_1)}{2\m^2}
=1+\frac{\cfrac{\m}{4}+\E_2}{\left|\cfrac{\m}{4}-\E_1\right|}
\rightarrow 1+\frac{\E_2}{\E_1}.
\end{eqnarray}
We realize that it is corresponds to the CM energy (\ref{Ecmx^+}) and (\ref{Ecmmaxparticle}) of the particle collision at $x=x^+_1$
with an equal mass $m\equiv m_1=m_2$ given by
\begin{eqnarray}
\frac{E^2_{\rm cm}(x^+_1)}{2m^2}=1+\frac{e_2}{e_1}.
\label{compare:EcmP}
\end{eqnarray}


\section{Discussion and Conclusion}
\label{Sec:Discussion and Conclusion}
We have considered the rear-end collision of two dust thin shells in the rotating BTZ spacetime to 
investigate the effects of the self-gravity of colliding objects on the high energy collision.
The shells divide the BTZ spacetime into three domains and the domains are matched by Darmois-Israel's method.
From the junction condition, all the domains must have the same angular momenta $J$. 
The angular momenta imply that the shells and domains corotate.

We have revealed that there are two effects of 
the self-gravity of thin shells.
First, we have shown that the mass of inner shell affects its critical condition~(\ref{critical:shell}).  
Second, the position of the event horizon changed from $x^H_1$ to $x^H_3$ because of the masses of two shells.

We have considered the shell collision in five cases according to the value of $J$.
The cases I~($J<\l M_1$), II~($J=\l M_1$), and III~($J=\l M_2$) would be especially interesting cases because of following reasons. 
The case I is a usual astrophysical situation as a black hole subextremely rotates and two objects collide near its event horizon.
In case II, the black hole extremely rotates initially 
and the collision of two falling shells corresponds with the BSW collision of two particles 
with an arbitrary high CM energy in the extremal black hole spacetime.
In case III, inner shell~1 can be satisfied the critical condition~(\ref{critical:shell}) 
that the effective potential for shell~$1$ becomes $V_1(x^H_2)=V_1'(x^H_2)=0$ 
on the extremal event horizon $x=x^H_2$ as with the BSW process in extremal black hole spacetimes~\cite{Banados:2009pr,Harada:2014vka}.
In cases I-III, the CM energy of the shells can be arbitrarily large 
if inner shell~$1$ satisfies the critical condition~(\ref{critical:shell}) and if outer shell $2$ does not.
However, an observer outside the event horizon $x^H_3$ cannot see the products of the collision with the arbitrary large CM energy
because it occurs inside the event horizon $x^H_3$.

If a shell collision occurs in a region $x \geq  x^H_3$, an observer who is outer than the collisional point may see products of the collision.
We have obtained the finite upperbound of the CM energy of the collision there.
Finally, we have concluded that the self-gravity of colliding objects suppresses its CM energy 
and the observer can only see the suppressed collision.

We have also found a test shell limit. 
We have shown that the CM energy and the effective potentials for shells in the test shell limit 
are very similar to the ones of particles.
The test shell limit would help us to understand the effect of the self gravity of the thin shells on the collisions.

We have considered only simple shell collisions on this paper.    
We hope that our paper stimulates further work on shell collisions 
and that researchers will investigate more realistic cases in the future.


\section*{Acknowledgements}
The authors would like to express the deepest appreciation to N.~Tanahashi for his insightful suggestions, comments, and discussions.
They thank also M.~Kimura, J.~V.~Rocha, T.~Harada, K.~i.~Nakao, O.~B.~Zaslavskii, A.~Naruko, T.~Kobayashi, T.~Kokubu, and F.~Hejda for valuable comments and discussions.
K.~O. was supported by JSPS KAKENHI Grant No.~18J10275.
They would also like to thank the Yukawa Institute for Theoretical Physics at Kyoto University, 
where this work was initiated during the YITP-X-16-10 on ``Workshop on gravity and cosmology for young researchers" 
supported by the MEXT KAKENHI Grant No.~15H05888.


\appendix
\section{
REAR-END COLLISION IN THE OVERSPINNING BTZ SPACETIME
}
\label{App:over-spinning_particle}
In this Appendix, we consider a rear-end collision of two particles with vanishing conserved momenta $L_1=L_2=0$ 
in the overspinning BTZ spacetime with $J>\l M$.

From Eq.~(\ref{V_particle}), the effective potential of particle $a$ takes the minimum value at $x=x^m$ as
\begin{eqnarray}
V_a(x^m)=-e^2_a+X,
\end{eqnarray}
where
\begin{eqnarray}
X\= f(x^m)=\frac{J}{\l}-M>0.
\end{eqnarray}
Particle $a$ with the nonpositive effective potential $V_a(x)$ can be within a region $x_a^-\leq x \leq x_a^+$
if a condition
\begin{eqnarray}
X\leq e_a^2,
\label{min:ea}
\end{eqnarray}
is satisfied.

The CM energy monotonically decreases from $E_{\rm cm}^{\rm max}$~(\ref{Ecmmaxparticle})
to 
\begin{eqnarray}
&&E_{\rm cm}(x^m)=\nonumber\\
&&\sqrt{m^2_1+m^2_2+2m_1m_2
\left[ \frac{e_1e_2-\sqrt{(e^2_1-X)(e^2_2-X)}}{X}\right] }, \nonumber\\
\end{eqnarray}
as $x$ increases from $x^-_a$ to $x^m$.
It monotonically increases 
and reaches $E_{\rm cm}^{\rm max}$ as $x$ increases from $x^m$ to $x^+_a$.


\end{document}